# Quantized dislocations for functional and quantum materials

Mingda Li[1] and Ricardo Pablo-Pedro[1]

[1]Department of Nuclear Science and Engineering
Massachusetts Institute of Technology, Cambridge, MA 02139

## Contents

## Abstract

*Dislocations have a profound influence on materials functional properties. In this perspective, we discuss the recent development of quantized dislocations – a theoretical tool that aims to compute the role of dislocations on materials' functionalities, at a full quantum field theoretical level. After a brief discussion of the motivation and a pedagogical introduction of quantization, we focus on a few case studies of dislon theory, to see how dislon can be applied to solve a given materials functionality problem and lead to new predictions. We conclude by visioning a few more open questions. With the aid of the powerful quantum field theory, the dislon approach may enable plenty opportunities to compute multiple functional and quantum properties in a dislocated crystal at a new level of clarity.*

## 1. Introduction

Dislocations affect materials mechanical properties and functionalities, such as electronic structure, optical properties, thermal transport, magnetic ordering, and superconductivity. For decades, the dislocation research has been centering on mechanical behaviors, such as crystal plasticity [5,6]. However, given the flourishing development of novel functional materials in recent years, the proper modeling of dislocation functionalities beyond mechanical aspects are gaining more and more importance, yet faces a series of challenges. As an extended defect with internal structure, a realistic dislocation is not only a quenched disorder composed of a distribution of strain field, but also subject to strong dynamical vibration and material-dependent Coulomb interaction in a complex interaction environment. In this sense, many dislocation studies are only describing a partial feature of a realistic dislocation, where its definition $\oint_L d\mathbf{u} = -\mathbf{b}$ is often not respected. Even all these factors are taken into account to fully characterize a realistic dislocation, the calculation on the functionalities poses another level of challenge. On the one hand, due to the long-range nature of a dislocation's stress field, the first-principles calculation with dislocations require a large supercell – could be as high as $N\sim1000$ atoms [7]. To carry out the response calculations, however, the computational complexity $\sim O(N^4)$ using the density functional perturbation theory is simply too high to be realistic [8]. On the other hand, since almost all those functionalities can be traced back to a microscopic quantum origin, a classical description of dislocation may simply be incapable to be integrated into a quantum theory in order to describe a complex quantum phenomenon.

In fact, a number of open questions remain in the field of dislocation functionalities. For electronic structure, it is known that the electrical resistivity in dislocated metals can have a particular type of electron-line defect resonance scattering, shown in Fig. 1a [9,10]. Although the resonance scattering model did explain some experimental data, the origin of such resonance is unclear: "*There must be some general mechanism underlying this phenomenon... This is still very much an open question...*"[9]. For optical properties, it is known that dislocation can induce luminescence [11] with 4 significant luminescence peaks (called D1 – D4) at low-temperature, which may survive even at room temperature after sample treatment (Fig. 1b). This phenomenon enabled dislocation based light-emitting diode (LED) applications [12]. However, even after three-decades-long research, the microscopic origin of these peaks is not fully understood until today, particularly "*The origin of the D2-line is still under discussion*" [13]. For the thermal transport, there has been a decades-long debate, arguing whether the dislocation-phonon interaction is static or dynamic in nature. Despite different temperature dependence, that $k_{\text{static}} \propto T^2$ while $k_{\text{dynamic}} \propto T^3$, carefully-planned experiments may exhibit a mixed behavior (Fig. 1c). In particular, one recent first-principles calculations of dislocations mentioned that "*Because of the breakdown of the Born approximation, earlier literature models fail, even qualitatively*"[2]. As to superconductivity, P.W. Anderson asserted that "*The (superconducting) transition temperature (with defects) will always be slightly smaller for the scattered states than they would be in the pure case*" [14], yet many experiments show otherwise. The anisotropic superconducting gap explains the transition temperature enhancement effect for point impurities [15], but a quantitative comparison for dislocations without using empirical parameter has been missing. These open questions, along with the unbearably high computational cost, indeed call for a much better approach to tackle the dislocation functionality problems at a more fundamental level.

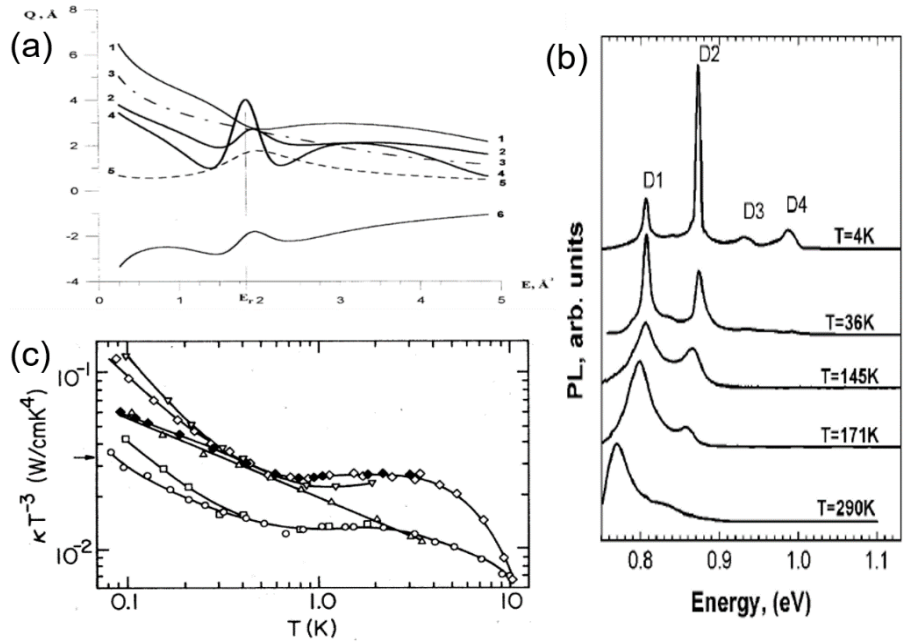

**Fig. 1.** (a) The computed cross section of electron-dislocation scattering, showing the resonance feature. (b) The temperature dependence of the dislocation luminescence in silicon. (c) The temperature-dependent thermal conductivity of prototypical dislocated LiF, showing a mixed T2 and T3 behavior. Figures are adapted from (Kveder et al. 2005; Roth and Anderson 1979).

This perspective article is a self-contained introduction on the recent theoretical progress of the so-called "dislon" theory. Dislon is a quasiparticle that aims to solve the above dislocation functionalities issues by directly quantizing a classical dislocation [1,3,4,16-19]. We believe that there are a number of unique advantages adopting this quantized dislocation approach, mainly formalism simplicity and strong predictive power.

*Formalism simplicity*: The simplicity has a multifold meaning. First is the procedure to incorporate a dislocation into an existing system. Apparently, there are plenty of existing approaches that can introduce a dislocation. For instance, for electronic property studies, a dislocation is often modeled as a line charge. With a scattering Coulomb potential $V(r)$, its Fourier transformation $V(q)$ gives the scattering strength, from which the electron-dislocation relaxation time can be obtained using Fermi's golden rule [20-22]. Taking the dislon approach, a dislocation is introduced by adding the dislon Hamiltonian $H_D$ and the interaction Hamiltonian $H_I$, into the original Hamiltonian $H_0$ for a pristine system. Given the arbitrary freedom to choose $H_0$, a dislocation can always be introduced properly. This leads to the second fold of simplicity that the Hamiltonian approach standardizes the procedure to compute all functional properties. With the knowledge of a given $H_0$, the electrical conductivity, thermal conductivity, thermopower, optical absorption, etc., are all computable using standard quantum many-body approaches, such as linear response theory [23]. Now the understanding of the role that dislocations may play is then reduced to a standard linear-response calculation of functionalities, but with the total Hamiltonian $H_0 + H_I + H_D$. The third meaning of simplicity lies in the form of dislon Hamiltonian $H_D$, which has great mathematical simplicity to tackle, but with all dislocation effects – strain, dynamic, Coulomb – incorporated simultaneously. This comprehensive description of dislocation is in sharp contrast with a potential scattering approach. Taking the popular line charge model of dislocation as an example, we see that the Burgers vector **b** does not even appear explicitly in the expression of relaxation time, which is unphysical to some extent.

*Strong predictive power*: Besides the formalism simplicity, the main advantage of dislon theory lies in the strong predictive power. This is a natural consequence by adopting a quantum field theoretical approach, since it can seamlessly incorporate all other interactions and correlation effects to an arbitrarily high order. The interaction effects are essential for a realistic scenario, for instance, dislocations may co-exist with point defects, phonons, and electrons in a real crystal, and we could always pick up the relevant degrees of freedom that are of interest for a given problem. For instance, if we are interested in phonon-dominant thermal transport, then the Coulomb charge model of dislocation cannot interact with phonon hence cannot be used for thermal transport study. The dislon theory, on the other hand, is a unified approach. Since it already contains the strain effect in the dislon field, the interaction with phonon becomes straightforward. The correlation effect and higher-order scattering, on the other hand, are both important toward qualitatively novel phenomena [24]. This also distinguishes the dislon theory from a semi-classical model. Again taking the line charge model as an example, we will know for sure that Coulomb scattering is guaranteed to happen, but meanwhile, full predictability is lost since there is no information how dynamic and strain effect affects the electronic structure. A quantum field theory, on the other hand, is still *ab initio* in nature, which retains a full predictive power without loss of information from the starting point.

## 2. Dislon as Quantized Dislocation

To see how a dislocation can be quantized into a quantized operator form, we noticed that both a dislocation and a phonon are atomic lattice displacement **u**, with the major difference come from the dislocation's topological constraint $\oint_L d\mathbf{u} = -\mathbf{b}$. Therefore, two major pillars are needed to quantize a dislocation. One is the lesson of quantum procedure, learned from the more familiar phonon quantization; the other is the unique features of a classical dislocation that distinguishes from a classical lattice wave.

*The lesson from phonon quantization:* We briefly outline the familiar phonon quantization first since the dislocation quantization shares some formalism similarity. A comprehensive procedure for phonon quantization can be found in quantum many-body monographs [23,25]. For a system with *N* atoms with mass *m*, each atom located at coordinate $\mathbf{R}_n$ has its own displacement $\mathbf{u}_n$ and momentum $\mathbf{p}_n$ ($n=1,2,...N$). The total kinetic energy gives

$$K = \sum_{n=1}^{N} \frac{|\mathbf{p}_n|^2}{2m} = \sum_{n=1}^{N}\sum_{i=1}^{3} \frac{(p_{ni})^2}{2m} \tag{1}$$

where $p_{ni}$ is the $i^{th}$ Cartesian component of the vector $\mathbf{p}_n$. The total potential energy can be written as

$$U = \frac{1}{2}\sum_{m,n=1}^{N}\sum_{i,j=1}^{3} u_{mi} V_{ij}^{mn} u_{ni} \tag{2}$$

where $u_{ni}$ is the $i^{th}$ Cartesian component of displacement $\mathbf{u}_n$, $V_{ij}^{mn}$ is a generalized "spring constant" for a harmonic oscillator. To obtain a quantum theory, we promote the classical dynamical variables $\mathbf{u}_n$ and $\mathbf{p}_n$ into first-quantized operators, with the following canonical quantization condition

$$[u_{ni}, p_{mj}] = i\hbar \delta_{ij} \delta_{mn} \tag{3}$$

i.e., different atomic locations ($m \neq n$) and directions ($i \neq j$) commute with each other.

As we see, the phonons in an *N*-atom system Eqs. (1) and (2) appear cumbersome in a first-quantized form. However, all these can be simplified using a second quantization approach. To do so, we perform a Fourier transform

$$\begin{aligned}\mathbf{u}_n = \frac{1}{\sqrt{N}}\sum_\mathbf{k}\mathbf{u}_\mathbf{k} e^{-i\mathbf{k}\cdot\mathbf{R}_n}, \; \mathbf{u}_\mathbf{k} = \frac{1}{\sqrt{N}}\sum_\mathbf{k}\mathbf{u}_n e^{i\mathbf{k}\cdot\mathbf{R}_n} \\ \mathbf{p}_n = \frac{1}{\sqrt{N}}\sum_\mathbf{k}\mathbf{p}_\mathbf{k} e^{-i\mathbf{k}\cdot\mathbf{R}_n}, \; \mathbf{p}_\mathbf{k} = \frac{1}{\sqrt{N}}\sum_\mathbf{k}\mathbf{p}_n e^{i\mathbf{k}\cdot\mathbf{R}_n}\end{aligned} \tag{4}$$

where $\mathbf{u}_\mathbf{k}$ and $\mathbf{p}_\mathbf{k}$ are the canonical displacement and momentum operator labeled by quantum number **k** (called crystal momentum), respectively. Then, substituting Eq. (4) back to Eqs. (1) and (2), and performing a Fourier transform to the expansion coefficient $V_{ij}^{mn}$,

$$D_{ij}(\mathbf{k}) = \sum_{\mathbf{R} \equiv \mathbf{R}_m - \mathbf{R}_n} V_{ij}^{mn} e^{-i\mathbf{k}\cdot(\mathbf{R}_m - \mathbf{R}_n)} \tag{5}$$

where since a crystalline solid is periodic with translation symmetry, we only need to sum over the position *difference* between two lattice positions $\mathbf{R}_n$ and $\mathbf{R}_m$. Now since $D_{ij}(\mathbf{k}) = D_{ji}(\mathbf{k})$, the $3\times 3$ $D(\mathbf{k})$ matrix ($\mathbf{k}$ can be considered as a parameter) can be diagonalized with real eigenvalues. The eigen-equation can be written as

$$D(\mathbf{k})\varepsilon_{\mathbf{k}\lambda} = m\omega_{\mathbf{k}\lambda}^2 \varepsilon_{\mathbf{k}\lambda} \tag{6}$$

in which $\varepsilon_{\mathbf{k}\lambda}$ is the eigenvector ($\lambda = 1,2,3$) called polarization vector, and $m\omega_{\mathbf{k}\lambda}^2$ is the eigenvalue. The second-quantized form can be defined by the particle occupation number formalism:

$$\begin{aligned}\mathbf{u}_{\mathbf{k}} &= \sum_{\lambda} \sqrt{\frac{\hbar}{2m\omega_{\mathbf{k}\lambda}}} \left(a_{\mathbf{k}\lambda}^+ + a_{\mathbf{k}\lambda}\right) \varepsilon_{\mathbf{k}\lambda} \\ \mathbf{p}_{\mathbf{k}} &= i\sum_{\lambda} \sqrt{\frac{\hbar m\omega_{\mathbf{k}\lambda}}{2}} \left(a_{\mathbf{k}\lambda}^+ - a_{\mathbf{k}\lambda}\right) \varepsilon_{\mathbf{k}\lambda}\end{aligned} \tag{7}$$

where $a_{\mathbf{k}\lambda}$ and $a_{\mathbf{k}\lambda}^+$ annihilate and create a phonon at state $\mathbf{k}\lambda$, respectively, satisfying the following commutation relation

$$[a_{\mathbf{k}\lambda}, a_{\mathbf{k}\lambda}^+] = 1 \tag{8}$$

Now substituting Eqs. (4), (7) and (8) back to Eqs. (1) and (2), we finally obtain a second-quantized Hamiltonian for 3D phonons:

$$H = \sum_{\mathbf{k}\lambda} \hbar\omega_{\mathbf{k}\lambda} \left(a_{\mathbf{k}\lambda}^+ a_{\mathbf{k}\lambda} + \frac{1}{2}\right) \tag{9}$$

which now has a much simpler form and enhanced power to deal with complex interaction problems. Such as formalism and power enhancement also happens to a classical dislocation upon second quantization.

*The lesson from Classical Dislocation*: Another major pillar that supports the quantized dislocation lies in a few aspects of its classical counterpart (Fig. 2). First, a dislocation exists in a crystalline solid, but not in amorphous materials (Fig. 2a). This greatly facilitates the electron and phonon interaction problems in that the Bloch's theorem is valid. Second, a dislocation not only can be defined in discrete crystals, but also can be defined in a continuous medium $\oint_L d\mathbf{u} = -\mathbf{b}$, where $\mathbf{u} = \mathbf{u}(\mathbf{R})$, and $\mathbf{R}$ is a continuous spatial coordinate (Fig. 2b). This facilitates a treatment to develop a simpler low-energy effective quantum theory. Third, a dislocation's definition $\oint_L d\mathbf{u} = -\mathbf{b}$ indicates a topological invariance and forbids a dislocation to end inside a crystal bulk (Fig. 2c). This facilitates the description of dislocation as a long straight line, without worrying a short or segmented dislocation line, if we leave the complex dislocation loop out of the picture.

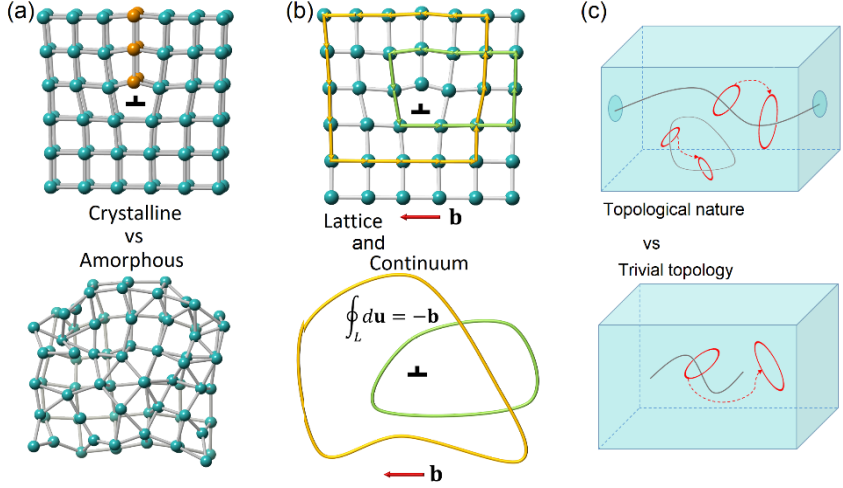

**Fig. 2.** A few essential elements for a classical dislocation, including crystalline, continuum compatible, and topological. Figures from [3].

With the method of canonical quantization procedure and the concept of topological invariance, the dislon theory can be considered as a natural merge of these two pillars. The detailed dislocation quantization procedure is introduced in a recent review [3]. Briefly speaking, a dislocated lattice system still contains kinetic energy and lattice strain potential energy, just like Eqs. (1) and (2). The dislocation's definition $\oint_L d\mathbf{u} = -\mathbf{b}$ is evolved into a simple boundary condition of the quantized operator. The incorporation of both the kinetic energy and potential energy is a critical move. This is so since many dislocation models contain only the potential energy part (e.g., stress field) in a frozen-lattice configuration. However, for phonon-dislocation scattering, it is well known that the dominant mechanism is the dynamic process, which could be orders-of-magnitude higher than the strain field [26,27]. By simultaneously incorporating both the kinetic energy and the potential energy, not only the formalism is greatly simplified in second-quantized form, but also both dynamic and static interactions are taken into account on an equal footing, which is more appropriate for a realistic dislocation.

## 2.1    The dislon Hamiltonian

The major difference between the phonon and dislocation quantization lies in the Fourier transform step. Since a dislocation is a localized 1D-like defect, we have another generic expansion instead of plane-wave expansion

$$\mathbf{u}_{dis}(\mathbf{R}) = \frac{1}{L^2}\sum_{\mathbf{k}} e^{i\mathbf{k}\cdot\mathbf{R}}\mathbf{F}(\mathbf{k})u_{\mathbf{k}} \qquad (10)$$

where $L$ denotes the system size, that there is one dislocation at present within a square of $L^2$. For acoustic phonons, we have $\mathbf{F}(\mathbf{k}) = \varepsilon_{\mathbf{k}}$. Moreover, under the long-wavelength limit $\mathbf{k} \to 0$, we have $\lim_{\mathbf{k} \to 0} u_\mathbf{k} = 0$ for phonons, i.e., it reduces to a perfect lattice under the static limit without displacement. The situation of dislocation is distinct. The $\mathbf{F}(\mathbf{k})$ is an expansion coefficient with local modes,

$$\mathbf{F}(\mathbf{k}) = \frac{1}{k_x k^2} \left( \mathbf{n}(\mathbf{b} \cdot \mathbf{k}) + \mathbf{b}(\mathbf{n} \cdot \mathbf{k}) - \frac{1}{(1-\nu)} \frac{\mathbf{k}(\mathbf{n} \cdot \mathbf{k})(\mathbf{b} \cdot \mathbf{k})}{k^2} \right) \quad (11)$$

if we assume $xz$ plane as slip-plane, and $\nu$ is the Poisson ratio.

On the other hand, $u_\mathbf{k}$ satisfies a different boundary condition $\lim_{k_z \to 0} u_\mathbf{k} = 1$ if we assume the dislocation is along the $z$-direction [17]. This simple boundary condition is a natural result to ensure the compatibility with the definition $\oint_L d\mathbf{u} = -\mathbf{b}$. In the end, the dislon Hamiltonian with dislon excitation $\Omega_\mathbf{k}$ in the second quantized form can be written as

$$H_D = \sum_{\mathbf{k} \geq 0} \Omega_\mathbf{k} \left( d_\mathbf{k}^+ d_\mathbf{k} + \frac{1}{2} \right) + \sum_{\mathbf{k} \geq 0} \Omega_\mathbf{k} \left( f_\mathbf{k}^+ f_\mathbf{k} + \frac{1}{2} \right) \quad (12)$$

where the operators satisfy

$$[d_\mathbf{k}, d_{\mathbf{k}'}^+] = \delta_{\mathbf{k}\mathbf{k}'}, \quad [f_\mathbf{k}, f_{\mathbf{k}'}^+] = \delta_{\mathbf{k}\mathbf{k}'} \quad (13)$$

with a boundary condition

$$\lim_{k_z \to 0} d_\mathbf{k} = \lim_{k_z \to 0} d_\mathbf{k}^+ = C_{k_\parallel} \quad (14)$$

in which $C_{k_\parallel}$ is a boundary term taking care of the effects along the in-plane directions perpendicular to the dislocation line, such as Coulomb scattering and strain field scattering. Whenever a classical effect needs to be taken into account, we can generalize $C_{k_\parallel}$ for a specific problem.

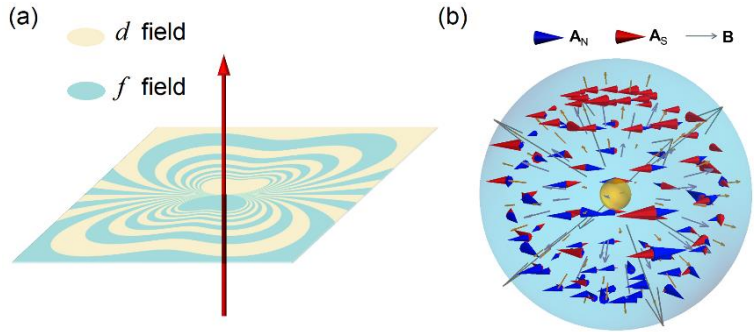

**Fig. 3.** (a) Two-fields of the dislon. (b) A magnetic monopole with two classical vector fields. Figure adapted [3].

To sum up, we need two Bosonic fields to describe a quantized dislocation, which are subject to different boundary conditions, in contrast to the phonon case where one Bosonic field suffices (**Fig. 3**a). This resembles another topological defect of magnetic monopole (**Fig. 3**b), where two classical fields (magnetic vector potential), $A_N$ and $A_S$, are needed to capture the intrinsic topology.

Eqs. (12) - (14) are the central results of a dislon Hamiltonian. Eq. (13) takes care of all dynamic effects, while Eq. (14) is responsible for the strain field scattering and the classical Coulomb scattering, where eventually $C_{k_\parallel}$ can be directly related to the classical dislocation-electron scattering amplitude.

## *2.2  General workflow to apply the dislon theory*

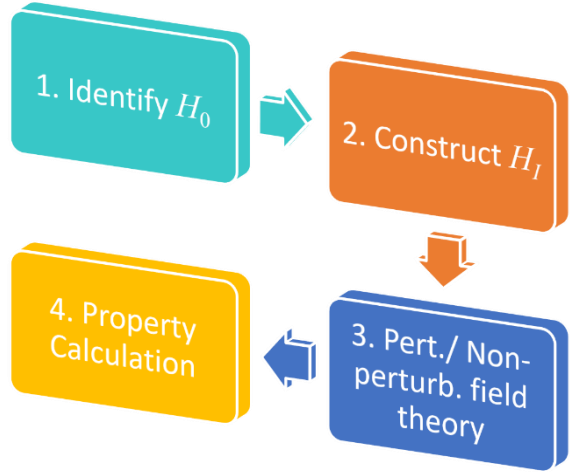

**Fig. 4.** The general workflow applying the dislon theory.

With the dislon Hamiltonian in hand, we are in good shape to introduce the general workflow applying the dislon theory to a general functionality problem, summarized in Fig. 4.

Step 1. Identify the Hamiltonian $H_0$ that describes the dislocation-free system. If we are interested in electrons in a metal, then $H_0$ is just non-interacting electrons; if we want to study the phonon transport, $H_0$ is the free-phonon Hamiltonian Eq. (9). If we want to study optical and magnetic properties that is influenced by dislocation, then a multi-band Hamiltonian can be applied, for instance, a 2-band model,

$$H_0 = \sum_{\mathbf{k}} \begin{pmatrix} c_{\mathbf{k}a}^+ & c_{\mathbf{k}b}^+ \end{pmatrix} \begin{pmatrix} E_{\mathbf{k}a} - \mu & 0 \\ 0 & E_{\mathbf{k}b} - \mu \end{pmatrix} \begin{pmatrix} c_{\mathbf{k}a} \\ c_{\mathbf{k}b} \end{pmatrix} \qquad (15)$$

where $(a,b) = (\uparrow, \downarrow)$ for a spin-1/2 system, while $(a,b) = (c,v)$ for a 2-band model with conduction and valence band.

Step 2. Identify the interacting Hamiltonian $H_I$. The non-interacting dislon Hamiltonian $H_D$ itself is not enough to affect $H_0$, without having an interaction term. As we will see in next Section, in many situations, we need to write down the classical interaction Hamiltonian and then perform the corresponding quantization. Fortunately, Since dislocation contains strain field, dynamic vibration and Coulomb interaction, the number of such interaction Hamiltonians are finite, some are quantized recently [17]. For instance, for the prototypical deformation potential scattering between dislocation displacement field and the gradient of Coulomb interaction, we have $\mathbf{u}_{dis} \cdot \nabla V_{Coulomb}$; for the velocity-velocity fluttering interaction with phonon, we have $\dot{\mathbf{u}}_{dis} \cdot \dot{\mathbf{u}}_{ph}$; while for the anharmonic dislocation-phonon interaction, we have $\mathbf{u}_{dis}\mathbf{u}_{ph}^2$, etc. The goal is to rewrite these classical interaction Hamiltonians in terms of a combination of creation and annihilation operators from $H_0$ (electron and phonon operators) and from $H_D$ (dislon operators).

Step 3. Perturbative and Non-perturbative calculations. With the non-interacting and interacting Hamiltonians, we are ready to carry out relevant calculations, by setting up perturbative and non-perturbative calculations. By non-perturbative, we refer to the functional integral approach [28], which is convenient to take into account the constraint Eq. (14). If, on the other hand, we only want to understand the dynamical quantum effect of dislocations, then no constraint is needed.

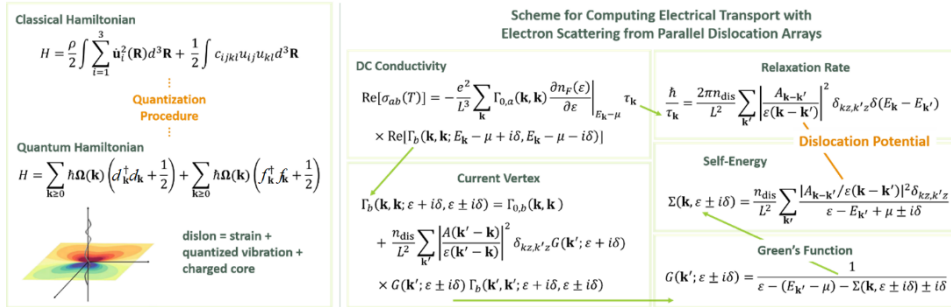

**Fig. 5.** The example workflow of electrical conductivity calculation using a quantum many-body approach with the presence of dislocations.

Step 4. Property calculation. Through step 3, almost all major physical quantities, such as DC and AC electrical conductivity, thermal conductivity, dielectric function, optical absorption, Seebeck coefficient, magnetic susceptibility, superconducting transition temperature, etc., can be computed systematically based on linear response theory, with controllable approximations. An example of DC conductivity workflow in shown in Fig. 5.

### 3. Cases Studies Using the Dislon Theory

This tribute intends to provide a few concrete examples and see how the dislon theory can be applied in a given interaction scenario. We will provide three examples, including the computation of the electron-dislon relaxation time, the calculation of superconducting transition temperature $T_c$, and also the phonon energy shift and relaxation time upon dislocation interaction. We elaborate the first example given its simplicity and outline the rest two by summarizing the main results and major predictions.

### *3.1 Computation of relaxation time*

In the classical dislocation theory, a dislocation is modeled as a scattering potential $V(\mathbf{r})$, and the relaxation rate from state *i* to *f* can be computed using the Fermi's golden rule:

$$\Gamma_{i \to f} = \frac{2\pi}{\hbar} \left| \int \psi_f^*(\mathbf{r}) V(\mathbf{r}) \psi_i(\mathbf{r}) d\mathbf{r} \right|^2 \delta(\varepsilon_f - \varepsilon_i) \quad (16)$$

where $\psi_i(\mathbf{r})$ and $\psi_f(\mathbf{r})$ are the initial and final wavefunctions with energy $\varepsilon_i$ and $\varepsilon_f$, respectively. If we assume a plane wave wavefunctions $\psi_i(\mathbf{r}) = e^{i\mathbf{k}\cdot\mathbf{r}}/L^{D/2}$ and $\psi_f(\mathbf{r}) = e^{i(\mathbf{k}+\mathbf{q})\cdot\mathbf{r}}/L^{D/2}$, and define a Fourier transform $V(\mathbf{q}) = \int V(\mathbf{r}) e^{-i\mathbf{q}\cdot\mathbf{r}} d^D\mathbf{r}$, the relaxation rate for an electron with momentum **k** is being scattered into **k+q** using Eq. (16) can be written as

$$\Gamma_{\mathbf{k} \to \mathbf{k}+\mathbf{q}} = \frac{1}{L^{2D}} \frac{2\pi}{\hbar} |V(\mathbf{q})|^2 \delta(\varepsilon(\mathbf{k}+\mathbf{q}) - \varepsilon(\mathbf{k})) \quad (17)$$

Since the momentum change **q** is arbitrary, if we assume a total number of $N_{dis}$ dislocations, the total relaxation rate as a function of electron momentum **k** can be written as

$$\Gamma_{\mathbf{k}} = N_{dis} \sum_{\mathbf{q}} \Gamma_{\mathbf{k} \to \mathbf{k}+\mathbf{q}} = \frac{2\pi}{\hbar} n_{dis} \int \frac{d^D \mathbf{q}}{(2\pi)^D} |V(\mathbf{q})|^2 \delta(\varepsilon(\mathbf{k}+\mathbf{q}) - \varepsilon(\mathbf{k})) \quad (18)$$

where $n_{dis}$ denotes the dislocation density.

Eq. (18) is a common approach to link dislocation's potential $V(\mathbf{r})$, electron energy $\varepsilon(\mathbf{k})$ to the relaxation rate $\Gamma_{\mathbf{k}}$. However, problems remain: a) The relation between $V(\mathbf{q})$ and the definition $\oint_L d\mathbf{u} = -\mathbf{b}$ is obscure. If we model a dislocation as a line charge, where $V(\mathbf{r})$ is Coulomb potential, it has nothing to do with dislocation's definition. b) When the electrons are under interactions in the solid, the wavefunctions are usually too difficult to obtain. In addition, it seems impossible to incorporate other interactions into this formalism. c) When dislocation density is high, multiple scattering emerges, it would be nice to have a formalism to treat multiple scattering, especially infinite order where a qualitative change of system such as Anderson localization, emerges [24]. d) Regardless of interaction and high-order effects, many other effects, such as temperature dependence, are also challenging to be incorporated. As we will see, the quantum field approach can solve all these problems instantly in an elegant way.

In a quantum field language, instead of using wavefunctions, the Green's functions, aka propagators, serve as building blocks. We will work with the imaginary-time Green's function

$$G(\mathbf{k}, \tau) = -\left\langle c_{\mathbf{k}}(\tau) c_{\mathbf{k}}^+(0) \right\rangle \quad (19)$$

which demonstrates a process that an electron with momentum **k** is created at time

0 ($c_\mathbf{k}^+(0)$ term), and then annihilated at time $\tau$ ($c_\mathbf{k}(\tau)$ term), hence represents the electron propagation probability amplitude under an interacting environment. $\tau$ is called imaginary time, $0 < \tau < \beta = 1/k_B T$, whose Fourier transform gives the so-called Matsubara frequency $p_n$, $p_n = (2n+1)\pi/\beta, n = 0, \pm 1, \pm 2...$ for fermions. Such formalism is a convenient approach to solve finite-temperature problems.

For non-interacting spinless electrons with $H_0 = \sum_\mathbf{k}(E_\mathbf{k} - \mu)c_\mathbf{k}^+ c_\mathbf{k}$, the non-interacting Green's function in Matsubara frequency domain is written as

$$G_0(\mathbf{k}, ip_n) = \frac{1}{ip_n - E_\mathbf{k} + \mu} \quad (20)$$

from which the most straightforward process that leads to relaxation time $\tau_\mathbf{k}$ can be considered as a virtual process where an electron $\mathbf{k}$ is first scattered into $\mathbf{k+q}$ by a dislon with momentum $\mathbf{q}$ with interaction strength $g_\mathbf{q}$, then the $\mathbf{k+q}$ electron is scattered back to the original momentum $\mathbf{k}$ through a dislon with momentum $-\mathbf{q}$. Such a process can be diagrammatically represented as:

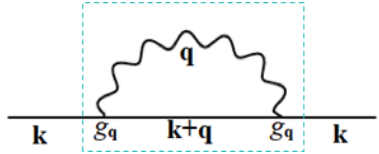

In above figure, the straight line segment and wavy line denotes the electron and dislon propagator, $G_0$ and $D_0$, respectively, in which the part inside the dashed rectangle is the lowest-order self-energy, which can be written as

$$\Sigma^1(\mathbf{k}, ip_n) = -\frac{1}{\beta}\sum_{\mathbf{q}m} g_\mathbf{q}^2 D^0(\mathbf{q}, i\omega_m) G^0(\mathbf{k-q}, ip_n - i\omega_m) \quad (21)$$

where the internal momentum $\mathbf{q}$ and Matsubara frequency $i\omega_m$ are summed over. Finally, the relaxation time $\tau_\mathbf{k}$ can be computed from a self-energy calculation:

$$\frac{1}{\tau_\mathbf{k}} = -2\,\text{Im}\left[\Sigma(\mathbf{k},0)\right] \quad (22)$$

which can lead to consistent result with Eq. (18). Despite seemingly more cumbersome than a Fermi's golden rule approach, the above quantum field scheme can indeed conquer all difficulties faced by Fermi's golden rule approach.

a) Dislocation's definition is well respected in the dislon Hamiltonian Eqs. (12) - (14).

b) If other interaction mechanisms are needed, no matter point defects or phonon interactions with electrons, or even Coulomb interaction between electrons, we only need to rewrite the corresponding electron propagator taking into account the relevant interactions. For instance, the lowest order correction of point defect or phonon scattering can be diagrammatically represented as:

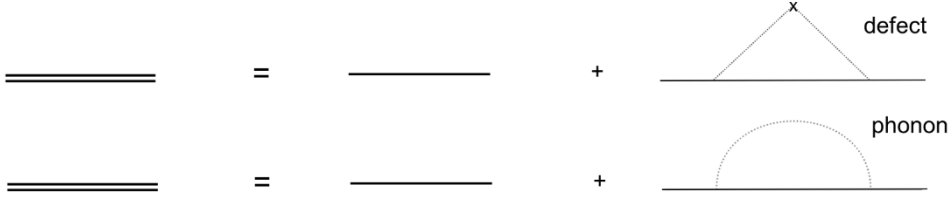

where the double-line on the left-hand side denotes the dressed electron propagator upon interaction. The total self-energy taking into account other interactions can be represented as:

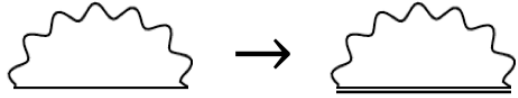

while the Eq. (22) is still valid by replacing the electron propagator $G_0 \to G$ in the updated self-energy.

c) Higher order scattering processes are easily incorporable as well. For instance, the second-order electron-dislon scattering can be written as pictorially as

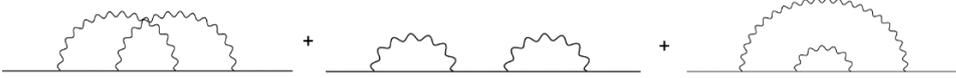

d) Other effects can also be incorporated in a systematic way. In particular, temperature effects are fully taken into account within the finite-temperature Matsubara formalism.

### *3.2  Electron-dislon interaction: Explaining $T_c$*

The power of adopting the dislon approach to study the electron-dislon interaction problem is beyond the calculation of relaxation time, but can also be used to study the effects of dislocations on dislocated superconductors. To see this, we notice that the electron-dislocation interaction Hamiltonian can well be described by a deformation potential scattering Hamiltonian

$$H_{e-ion} = \int d^3\mathbf{R}\rho_e(\mathbf{R})\sum_{n=1}^{N}\nabla_{\mathbf{R}}V_{ei}(\mathbf{R}-\mathbf{R}_n^0)\cdot \mathbf{u}_{dis}(\mathbf{R}_n^0) \qquad (23)$$

where $\rho_e(\mathbf{R})$ is the electron charge density, $V_{ei}$ is the Coulomb interaction, $\mathbf{R}_n^0$ is the atomic location of atom number $n$. After quantization, we have

$$H_{e-dis} = \sum_{\mathbf{k}'\sigma;\mathbf{k}\geq 0} g_{\mathbf{k}} c^+_{\mathbf{k}'+\mathbf{k}\sigma}c_{\mathbf{k}'\sigma}d_{\mathbf{k}} + \text{h.c.} \qquad (24)$$

where $g_{\mathbf{k}} \propto \mathbf{k}\cdot \mathbf{F}(\mathbf{k}) \propto \dfrac{1-2\nu}{1-\nu}b$ is the electron-dislocation coupling strength, which gives highly consistent result from semi-classical theory [29].

In the end, we were able to compute the superconducting transition temperature $T_c$, quantitatively in a dislocated superconductor, for the first time. The $T_c$ equation can be written as [4]

$$\frac{1}{g_{ph}+g_D} = N(\mu)\int_0^{\theta_D'} d\xi \sum_{s=\pm 1} \tanh\left(\frac{\xi+is\Gamma_D}{2T_c}\right)\bigg/2\xi \quad (25)$$

with two major coefficients (classical potential scattering $\Gamma_D$ and quantum fluctuation $g_D$)

$$\Gamma_D = \frac{\pi m^*}{4\hbar^2 k_F^2 k_{TF}^4}\left(Ze^2 n\right)^2 n_{dis} b^2 \left(\frac{1-2\nu}{1-\nu}\right)^2$$

$$g_D = n_{dis}^{3/2} \frac{\left(4\pi Ze^2 n\right)^2}{2k_{TF}^4(\lambda+2\mu)} \quad (26)$$

where $g_{ph}$ is the electron-phonon coupling constant, $\theta_D'$ is the renormalized Debye frequency, $m^*$ is the effective mass, $k_F$ is Fermi wavevector, $k_{TF}$ is Thomas-Fermi screening wavevector, $n$ is atomic number density, $n_{dis}$ is dislocation density, $b$ is Burgers vector, $\nu$ is Poisson ratio, $\lambda$ and $\mu$ are Lame parameters, which are all common materials parameters that can be looked up in database. The Eq. (25) implies a completion effect between classical scattering and quantum fluctuation, as illustrated in Fig. 6, which shows excellent quantitative agreement comparing with experimental data without free fitting parameter [4].

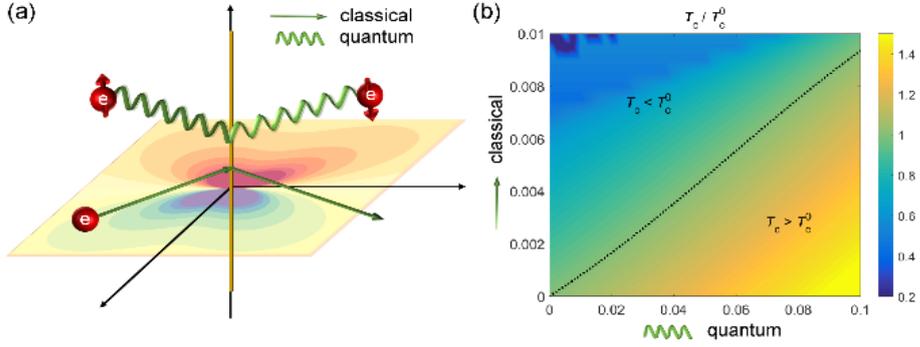

**Fig. 6.** (a) Schematics of the classical vs quantum type of electron-dislocation interaction, whose competition will determine the $T_c$ in a dislocated superconductor (b). Figure adapted from [4].

### 3.3 *Phonon-dislon interaction*: *Beyond perturbation*

The general workflow is also applicable to the dislocation-phonon interaction problems. The classical velocity-velocity drag-like fluttering interaction Hamiltonian between dislocation and phonon can be written as

$$H_{flu} = \rho\int \dot{\mathbf{u}}_{ph}(\mathbf{R})\cdot\dot{\mathbf{u}}_{dis}(\mathbf{R})d^3\mathbf{R} \quad (27)$$

After the second quantization procedure, we have

$$H_{flu} = \sum_{\mathbf{k}>0} g_{\mathbf{k}}^{dyn} \boldsymbol{\varepsilon}_{\mathbf{k}} \cdot \mathbf{F}(\mathbf{k})\left(-a_{\mathbf{k}} + a_{-\mathbf{k}}^{+}\right) f_{\mathbf{k}} + \text{h.c.} \qquad (28)$$

where $g_{\mathbf{k}}^{dyn}$ is the dislocation-phonon coupling dynamic coefficient that depends on phonon and dislon dispersions $\omega_{\mathbf{k}}$ and $\Omega_{\mathbf{k}}$. The interaction Hamiltonian Eq. (28) turns out to be able to shift phonon dispersion and meanwhile result in a finite phonon lifetime [1].

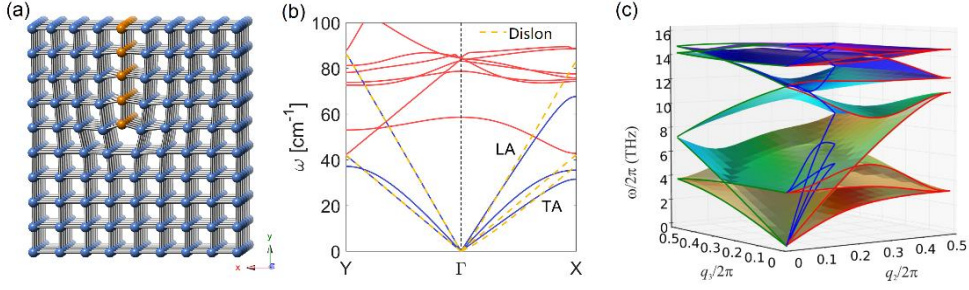

**Fig. 7.** (a) A schematic of a dislocated simple cubic crystal. (b) An isotropic phonon softening and transverse acoustic (TA) modes splitting with dislocations. (c) Verification through *ab initio* calculations. Figures adapted from [1,2].

The phonon dispersion changed by dislocation is shown in Fig. 7. In a dislocated crystal (Fig. 7a), the dislon theory predicts anisotropic phonon softening and TA modes splitting (Fig. 7b, yellow lines), which are confirmed by both lattice dynamics simulations (Fig. 7b, blue lines) and *ab initio* first-principles calculations (Fig. 7c). As to the finite phonon lifetime, the dislon theory reveals a fruitful structure, depending on the dislocation type, phonon type and phonon polarization.

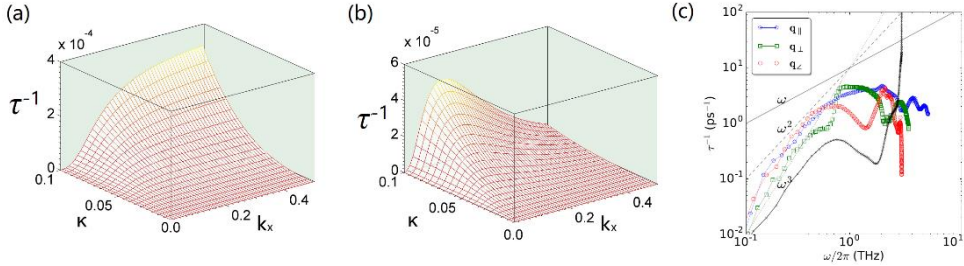

**Fig. 8.** The dislocation-phonon scattering relaxation rates for a transverse-like (a) and a longitudinal-like (b) with an edge dislocation. A similar saturation behavior in (a) and a resonance behavior in (b) are seen in the *ab initio* calculations in (c). Figures adapted from [1,2].

In particular, a special type of saturation and resonance are predicted (Fig. 8a, b). These go far beyond any classical theory, where the relaxation rate $\tau^{-1}$ only contains monotonic dependence with phonon frequency ω:

$$\tau_{\text{fluttering}}^{-1} \propto \omega^{-1}$$
$$\tau_{\text{strain}}^{-1} \propto \omega$$
$$\tau_{\text{core}}^{-1} \propto \omega^3 \qquad (29)$$
$$\tau_{\text{isotope}}^{-1} \propto \omega^4$$

where the four terms denote phonon-dislocation dynamic "fluttering" scattering, phonon-dislocation strain-field scattering, phonon-dislocation core scattering, and phonon –isotope point defect scattering, respectively. Both the saturation behavior and the resonance behavior are revealed in the *ab initio* calculations (Fig. 8c), although the simulation and theory are not precisely computing the same system. The simulation contains a dislocation pair to reduce the system size, while the dislon theory is computed in a simpler system with single dislocation, thus has a slightly different definition on the phonon modes.

## 4.     Outlook and Perspective

Despite some initial success of the dislon theory, as a framework on which new theories can be built upon, the dislon theory is still at its infant stage. Here we sketch a few more examples that appear challenging for a classical dislocation theory, but are expected to be directly computable using the dislon theory:

*Electronic bandgap*: how dislocations can change the bandgap in semiconductors and insulators. We notice that the multi-band Hamiltonian Eq. (15) contains an energy gap. By introducing an interaction Hamiltonian, such as Eq. (24), the bandstructures will change, $E_{\mathbf{k}a} \to E'_{\mathbf{k}a}$, which can be obtained from perturbation theory, and will result in a bandgap modulation effect.

*Optical absorption*: the role of dislocations to the optical absorption coefficient $\alpha(\omega)$. Optical absorption leads to intensity distinction $I(x) = I_0 \exp(-\alpha x)$. To compute $\alpha(\omega)$, we first follow the scheme like Fig. 5 to compute the AC conductivity $\sigma(\omega)$, using the original Hamiltonian $H_0$ and total Hamiltonian $H_0 + H_I + H_D$. Then, the dielectric function $\varepsilon(\omega)$ can be written as $\varepsilon(\omega) = 1 + 4\pi i \sigma(\omega)/\omega$. Finally, defining an extinction coefficient $k(\omega)$ as $k^2(\omega) \equiv (|\varepsilon(\omega)| - \text{Re}[\varepsilon(\omega)])/2$, the optical absorption coefficient can be written as $\alpha(\omega) = 2\omega k(\omega)/c$.

*Phase transition*: Although challenging to summarize within a few steps, one major application that distinguishes the dislon theory from a classical dislocation model lies in the capability to study dislocation-induced phase transition problems, such as metal-insulator transition. Early experiments reported the dislocation-induced semiconductor-superconductor phase transition [30], which has been explained using the dislon theory, quantitatively [4]. In a more recent example, a quantum phase transition of phonons is revealed [19], caused by a competition effect between the topological protection of dislocation and a topological-breaking inelastic scattering. By understanding the relationship between dislocations and phase transitions, dislocations will gain more importance from materials imperfections to a new dimension to tailor the phase diagram, including unexplored new electronic, phononic, and photonic phases.

Indeed, the dislon theory offers a systematic approach to understand the interplay between crystal dislocations and materials electronic, spintronic, phononic and photonic degrees of freedom at a microscopic quantum level. However, why the dislon, or say the quantization procedure, could work in the first place? Intuitively, the term "quantum field" contains both the "quantum" part, which deals with the internal dynamics and excitation, and the "field" part, which deals with spatial extension. This dual-nature seems appropriate to describe a dislocation, which contains both internal dynamics, such as dynamic fluttering, and spatial extension that arises naturally as an extended defect. In fact, the spatial extension naturally and unavoidably leads to the internal dynamic structure. This brings up a more general question, that whether any extended defect can, or even should, be described by some quantum field:

$$\text{Large Defects} \;=\; \text{Quantum Fields} \qquad (30)$$

The answer to this grand question may greatly empower our approach to tackle with complex defects problems.

**References**

[1] M. Li, Z. Ding, Q. Meng, J. Zhou, Y. Zhu, H. Liu, M. S. Dresselhaus, and G. Chen, Nano Lett **17**, 1587 (2017).
[2] T. Wang, J. Carrete, A. van Roekeghem, N. Mingo, and G. K. H. Madsen, Physical Review B **95**, 245304 (2017).
[3] M. Li, in *ArXiv e-prints(*2018), p. 1808.07777.
[4] M. Li, Q. Song, T.-H. Liu, L. Meroueh, G. D. Mahan, M. S. Dresselhaus, and G. Chen, Nano Lett **17**, 4604 (2017).
[5] F. R. N. Nabarro, *Theory of crystal dislocations* (Oxford, Clarendon P., 1967., 1967), International series of monographs on physics.
[6] J. P. Hirth and J. Lothe, *Theory of dislocations* (Krieger Pub. Co., Malabar, FL, 1992), 2nd edn.
[7] Z. Wang, M. Saito, K. P. McKenna, and Y. Ikuhara, Nat Commun **5**, 3239 (2014).
[8] S. Baroni, S. de Gironcoli, A. Dal Corso, and P. Giannozzi, Rev Mod Phys **73**, 515 (2001).
[9] R. A. Brown, J Phys F Met Phys **7**, 1283 (1977).
[10] A. S. Karolik and A. A. Luhvich, J Phys-Condens Mat **6**, 873 (1994).
[11] N. A. Drozdov, A. A. Patrin, and V. D. Tkachev, ZhETF Pisma Redaktsiiu **23**, 651 (1976).
[12] V. Kveder, M. Badylevich, E. Steinman, A. Izotov, M. Seibt, and W. Schroter, Appl Phys Lett **84**, 2106 (2004).
[13] M. Reiche and M. Kittler, Crystals **6**, 74 (2016).
[14] P. W. Anderson, J Phys Chem Solids **11**, 26 (1959).
[15] D. Markowitz and L. P. Kadanoff, Phys Rev **131**, 563 (1963).
[16] M. Li, W. Cui, M. S. Dresselhaus, and G. Chen, New J Phys **19**, 013033 (2017).
[17] M. Li, Y. Tsurimaki, Q. Meng, N. Andrejevic, Y. Zhu, G. D. Mahan, and G. Chen, New J Phys **20**, 023010 (2018).

[18]	C. Fu and M. Li, J Phys-Condens Mat **29**, 325702 (2017).
[19]	R. Pablo-Pedro, N. Andrejevic, Y. Tsurimaki, Z. Ding, T.-H. Liu, G. D. Mahan, S. Huang, and M. Li, in *ArXiv e-prints(*2018), p. 1809.06495.
[20]	D. Jena, A. C. Gossard, and U. K. Mishra, Appl Phys Lett **76**, 1707 (2000).
[21]	D. C. Look and J. R. Sizelove, Phys Rev Lett **82**, 1237 (1999).
[22]	N. G. Weimann, L. F. Eastman, D. Doppalapudi, H. M. Ng, and T. D. Moustakas, Journal of Applied Physics **83**, 3656 (1998).
[23]	G. D. Mahan, *Many-particle physics* (Kluwer Academic/Plenum Publishers, New York, 2000), 3rd edn., Physics of solids and liquids.
[24]	J. Rammer, *Quantum transport theory* (Perseus Books, Reading, Mass., 1998), Frontiers in physics, 99.
[25]	A. L. Fetter and J. D. Walecka, *Quantum theory of many-particle systems* (Dover Publications, Mineola, N.Y., 2003).
[26]	A. Maurel, J.-F. Mercier, and F. Lund, Physical Review B **70** (2004).
[27]	A. C. Anderson and M. E. Malinowski, Physical Review B **5**, 3199 (1972).
[28]	J. W. Negele and H. Orland, *Quantum many-particle systems* (Addison-Wesley Pub. Co., Redwood City, Calif., 1988), Frontiers in physics, 68.
[29]	D. L. Dexter and F. Seitz, Phys Rev **86**, 964 (1952).
[30]	N. Y. Fogel, A. S. Pokhila, Y. V. Bomze, A. Y. Sipatov, A. I. Fedorenko, and R. I. Shekhter, Phys Rev Lett **86**, 512 (2001).